\documentstyle[preprint,aps,psfig]{revtex}

\begin{document}
\preprint{SPbU--IP--97--08}
\title{On the stability problem in  the
$O(N) $ nonlinear sigma model. }
\author{S. \'{E}.  Derkachov }
\address{
 Department of Mathematics,
St.-Petersburg Technology Institute, St.-Petersburg, Russia \\
}
\author{A. N. Manashov}
\address{
Department of Theoretical Physics,  Sankt-Petersburg State
University, St.-Petersburg, Russia
\\}
\date{\today}
\maketitle
\begin{abstract}
The stability problem for the
$ O(N) $ nonlinear sigma model in the
$ 2+\epsilon $  dimensions is considered. We present the results
of the
$ 1/N^{2} $   order calculations of the critical exponents (in the
$ 2<d<4 $ dimensions)  of the composite operators relevant for this
problem. The arguments in the  favor of the scenario
with the conventional fixed point are given.
\end{abstract}


 In a recent time the possibility of a scenario with the
nonconventional fixed points in
$ 2+\epsilon  $ expansions for various models was widely discussed. It
has been  firstly observed by Kravtsov, Lerner and Yudson in
$ Q $ matrix model~\cite{KLY}, and proved later by Wegner for
$ N $ vector, unitary and orthogonal matrix models~\cite{WZ,W1,W2},
that a certain class of composite operators with
$ 2s $ fields and
gradients endangers the stability of nontrivial fixed points in
$ 2+\epsilon $ expansions for these models.  In the following we
restrict ourselves to the case of the
$ N $ vector model.  The one~-~loop critical exponents of the corresponding
operators in
this model read as~\cite{WZ}
\FL
\begin{equation}
x_{s}=2s-\epsilon\frac{s(s-1)}{N-2}+O(\epsilon^{2}).
\label{ex2plus}
\end{equation}
 Judging from these results one could argue
that for a sufficiently large
$ s $ operators  become relevant
(operators with
$ y>0 $,
$ y=0 $ and
$ y<0 $  ($ y\equiv d-x $) are relevant, marginal and irrelevant,
resp.)
and the conventional nontrivial fixed point
becomes unstable against an infinite number of high--gradient perturbations.
This would have important consequences for the present understanding
of many problems that rely on $2+\epsilon$~expansions.

To get a more detailed notion of this problem it seems a reasonable to
take advantages of the
$ 1/N $ expansions, which being nonperturbative in its nature
relates the
$ 2+\epsilon $ and
$ 4-\epsilon $ expansions. The latter is commonly believed to be free
of any pathologies.
Moreover, in this approach
the corresponding operators look extremely
simple~---~$ \sigma^{s} $, where
$ \sigma $ is the auxiliary field.
The critical exponents for this set of operators
had been calculated in Ref.~\cite{Step}, and
independently in Ref.~\cite{Ruhl}:
\FL
\begin{equation} \label{VSr}
x_{s}=2s+\frac{\eta_{1}}{N}s(d-1)
\frac{(s-1)d(d-3)-2(d-2)}{4-d},
\end{equation}
where
{$\eta_{1}=4(2-\!\mu){\Gamma(2\mu-\!2)}/{\Gamma^{2}(\mu-\!1)\Gamma(2-\mu)}
{\Gamma(\mu+1)}$} and
$ \mu\equiv d/2 $. One can see that the problem similar to those in
$ 2+\epsilon $ expansion arises below
$ d=3 $.

The next attempt to attack  the stability problem
has been undertaken in the papers~\cite{CC,CC1},
where the critical exponents~
$ x_{s} $ have been calculated with the
$ \epsilon^{2} $ accuracy.
However, the full answer for~$ x_{s} $ given in Refs.~\cite{CC,CC1}
is not consistent with those~(Eq.~(\ref{VSr}))
obtained in Refs.~\cite{Step,Ruhl} in the
$ 1/N $ approach,
and with the expression
  for  index
$ \omega $  ($ \omega=x_{2}-d $), recently obtained in
Ref.~\cite{jag} with
$ 1/N^{2} $  accuracy.

The aim of the present paper is the calculation of the
critical exponents
$ x_{s} $ with $ 1/N^{2} $ accuracy.
This provides us the deeper insight into problem and allow
to suggest the realistic scenario for its solution.
The computation scheme developed below is of own interest.
We remind that only  few critical indices are know beyond the
$ 1/N $  order~\cite{VPH,VPH2,jag,jag2}.
All of them had been calculated by the
specific methods, which are not applicable in a general situation.
With this remark
we postpone the discussion to the end of the paper and proceed to the
calculations.
\vskip 0.5cm

The renormalized action for the nonlinear sigma model can be written
as~\cite{Nal}:
\FL
\begin{equation}
S = -\frac{Z_{1}}{2}(\partial\phi)^{2}
 -\frac{1}{2}\sigma K_{\Delta}{M^{-2\Delta}} \sigma
 +\frac{Z_{2}}{2}\sigma
\phi^{2}
 +\frac{1}{2}\sigma K \sigma.
\label{action}
\end{equation}
Here
$ \phi^{A} $ is the vector field ($ A=1,\ldots,N $),
$ \sigma $ is the auxiliary scalar field;
$\Delta$
and
$ M $
are regularization parameter and
mass, respectively.
The kernel
$ K $
is determined from the requirement of the cancellation
of the self--energy diagram contribution of the sigma field:\
$ K(x)=-N/2G_{\phi}^{2}(x) $.
$ G_{\phi}(x) $
is the propagator of
the $ \phi^{A} $
field
($<\phi^{A}(x)\phi^{B}(0)>=\delta^{AB}G_{\phi}(x)$).
The regularized kernel
$ K_{\Delta} $
is defined as
$K_{\Delta}(x)=K(x)x^{-2\Delta}$.
The divergencies appearing at the calculations of the Feynman
diagrams as poles in
$ \Delta $ are removed by the suitable choice of the constants
$Z_{1}$ and
$ Z_{2} $. Henceforth we will use the minimal subtraction
scheme (MS).


Unfortunately, the model under consideration is
not the multiplicatively renormalized~\cite{Nal}. This means that
the freedom in the choice of the renormalization prescriptions can
not be compensated by the redefinition of the fields and  the
parameters of  Lagrangian. Thus one fails to apply the standard RG methods
for the calculations of the critical indices. The general discussion
of this topic can be found in Refs.~\cite{Nal,Step}. Here we
remind only one basic property  important for the following,
namely, the correlation functions of the fields and composite
operators in the model under consideration are scale
invariant.
There is,  the renormalized one particle irreducible
Green's function
$ \Gamma^{O,n}(p,p_{i})=\langle O(p)\Phi(p_{1})...\Phi(p_{n})\rangle$
with  operator insertion ($ \Phi\equiv \phi, \sigma $)
satisfies the equation
$\Gamma^{O,n}(\lambda p,\lambda p_{i})= \lambda ^{{x_{O}
-n_{\Phi}{x}_{\Phi}}}\Gamma^{O,n}(p,p_{i}) $.
As usual,
$n_{\Phi}x_{\Phi}$ means
$
 n_{\phi}x_{\phi}+n_{\sigma}x_{\sigma}$ and
$ x_{O}, x_{\Phi} $ are the critical dimensions of operator
$ O $ and fields
$ \Phi $, respectively.
In the following the composite operators and Green functions
are assumed to be
renormalized, i.e. the all needed counterterms are taking into
account.

 The effective algorithm for the computation of
the anomalous dimensions of composite operators in the first
order of $ 1/N $ expansion exploiting the property
of scale invariance
has been
developed in the paper of Vasil'ev
and Stepanenko~\cite{Step}. They have shown that the
simple correlation between the $ \Delta $ pole residues of the
Green's functions and the corresponding anomalous dimensions
exists.
The generalization of this method to all order
of
$ 1/N $ expansion (VS scheme) is given below.
The basic formula  we used for the calculation of the anomalous
dimensions reads:
\begin{equation}
  u\Gamma^{O,n}(p,p_{i},M)=\lim_{\Delta\to 0}2
\Delta\sum_{\{ G\}}
  n_{\sigma}^{G}\Gamma_{G}^{O,n}(p,p_{i},M,\Delta).
\label{main}
\end{equation}
Here
$  \Gamma^{O,n}(p,p_{i},M) $ is the
$ 1PI $ $n$~--~point Green's  function with operator insertion.
A operator
$ O $ is assumed to be the operator with dimension (scaling one).
The sum runs over whole set of the diagrams
(including those with counterterms);
$ n_{\sigma}^{G} $ is  the number of the sigma lines in the
diagram $ G $;
$ u=-\gamma_{O}+n_{\Phi}\gamma_{\Phi} $
($ \gamma $ is used for  anomalous dimensions).

To derive the Eq.~(\ref{main}) we note that the scale invariance
results in the following form of $  \Gamma^{O,n}(p,p_{i},M) $:
\begin{equation}
 \Gamma^{O,n}(p,p_{i},M)
\equiv\lim_{\Delta\to 0}
\sum_{\{ G\}}
  \Gamma_{G}^{O,n}(p,p_{i},M,\Delta)
=p^{U}(M/p)^{u}\widetilde\Gamma(p_{i}/p)
\label{qq}
\end{equation}
where
$ U=x^{can}_{O}-n_{\Phi}x^{can}_{\Phi} $.
Then
acting by $ M\partial_{M} $ on
 $\Gamma^{O,n}(p,p_{i},M) $
 and taking into account that
the only dependence on
$ M $ in the diagrams results from
the propagator of
$ \sigma $  field
($ G_{\sigma}=M^{2\Delta}K_{\Delta}^{-1} $)
(we remind that all counterterms are chosen independent on
$ M $ (MS scheme))
 one immediately obtains the Eq.~(\ref{main}).
Note,  the finiteness of the lhs
of Eq.~(\ref{main}) ensures the cancellation of all
$ \Delta $ poles in the rhs except for the first order ones.

 Further, in the case of  operators
$ \{ O_{i}\}$ mixing under
renormalization  one should seek for the proper scaling
operators as the linear combinations:
$ {\tilde O}_{i}=\sum c_{ik}O_{k} $.
To determine both the anomalous dimensions and the form of those
one should consider the
 Eq.~(\ref{main})
for
$ n $~--~point  Green's functions with insertion of operators
$ {\tilde O}_{i} $ for different
$ n $.
\vskip 0.3cm

We apply now the above scheme to the calculation of the critical
exponents of
$ \sigma^{s} $ operators in
$ 1/N^{2} $ order. In spite of the fact that there are a lot of
operators with the same canonical dimension, which could have
admixed  to
$ \sigma^{s} $, this does not happen in the first order of
$ 1/N $ expansion~\cite{Step}.  The renormalized
$ \sigma^{s} $ operator reads as
 $ [\sigma^{s}]=Z_{s}\sigma^{s} $
($ Z_{s}=1+q_{s}/N\Delta $)  and is the proper scaling operator in this
order. (Henceforth we use the standard notation
$ [O] $ for the renormalized operator.) Going to the next order
one finds  the counterterms of the following form are
required~--~($ \sigma^{s},\> \sigma^{s-2}\partial^{2}\sigma,\>
\partial^{2}\sigma^{s-1} $), while all diagrams describing
$ \sigma^{2}\to \partial^{4} $ transition have not the divergencies.
This force us to conclude that in
$ 1/N^{2} $ order the scaling operators under consideration have form
\FL
\begin{equation}
 [O_{s}]=[\sigma^{s}]+\frac{\alpha_{1}}{N}
[\sigma^{s-2}\partial^{2}\sigma]+
\frac{\alpha_{2}}{N}[\partial^{2}\sigma^{s-1}] +O(\frac{1}{N^{2}}).
\label{Os}
\end{equation}
To determine
$ \alpha_{1,2} $ (really, we need
$ \alpha_{1} $ only) one must consider Eq.~(\ref{main}) for
$ (s-1) $ point Green's
function of sigma fields with insertion of the operator
$ [O_{s}] $.
Doing the same for
$ s $~--~point function
$ \Gamma^{O_{s},s}(p, p_{i})$ one obtains
the anomalous dimension of the operator in hand.

However, there is the much simple and elegant way to solve
the mixing problem.
Let us remind that the nonlinear
$ \sigma $ model is the simplest example of the conformal field
theory (CFT) ($ d>2 $)~\cite{Ruhl,Osborn,Petkou}.
Following along the
lines of the paper~\cite{Pol} we suppose that  the conformal operators
(CO's)
and
their total derivatives form a complete basis in the space of all
operators. In this case
any exact scaling operator
not being a total derivative of other is a conformal one, the
opposite is evidently true.
 (Of course, the more accurate statement needs when there is a
degeneracy of critical dimensions.)
 This observation being combined
with other the well known fact~--~vanishing of two~--~point correlator
of CO's with different scaling
dimensions~\cite{Pol}~--~considerably simplifies
the solution of the mixing problem.
For the conciseness we illustrate this idea on the concrete examples,
the
generalization being straightforward.

The only nonderivative scaling (and hence  conformal)
operators on the levels
$ s=1,2 $  (i.e. with the canonical dimensions equal to
$2s $)
 have form
$ [O_{1}]=[\sigma] $  and
$ [O_{2}]=[\sigma^{2}]+\alpha(N)\partial^{2}\sigma $.
From the requirement of "orthogonality"
$ \langle O_{2}(x)O_{1}(y) \rangle =0 $ one immediately obtains
$\langle [\sigma^{2}(p)]\sigma(-p) \rangle_{1PI}=\alpha(N)p^{2} $,
with $ \alpha(N) =(\eta_{1}/4N)\>(\mu-1)(2\mu-3)/(3-\mu)+O(1/N^{2}) $.
The above equality holds in
all order of $ 1/N $ expansion.
Note, namely the absence of the logarithmic
corrections to the above correlator results in the multiplicative
renormalization of
$ \sigma^{s} $ operators in the first
$ 1/N $ order~\cite{Step}.

On the level
$ s=3 $ two new conformal operators come into a game. On the
classical level they reads as \mbox{$O^{(1)}_{3}=\sigma^{3}$} and $
O^{(2)}_{3}=\sigma\partial^{2}\sigma- {(3-\mu)\over{2(5-\mu)} }
\partial^{2}\sigma^{2} $.
Then the exact CO's can be written as:
\FL
\begin{mathletters}
\label{example}
\begin{eqnarray}
{[O_{3}^{c,(1)}]}=[O^{(1)}_{3}]+a_{1}[O^{(2)}_{3}]+a_{2}\partial^{2}
[O_{2}]+a_{3}\partial^{4}[O_{1}]
\label{example:1}\\
{[O_{3}^{c,(2)}]}=[O^{(2)}_{3}]+b_{1}[O^{(1)}_{3}]+b_{2}\partial^{2}
[O_{2}]+b_{3}\partial^{4}[O_{1}]
\label{example:2}
\end{eqnarray}
\end{mathletters}
Here
$ a_{i}, b_{i} $  are some functions of
$ N $ and
$ \mu $. From the "orthogonality" of
$[O_{3}^{c,(1,2)}] $  to
$ [O_{2}],[O_{1}] $  one easy finds that
$a_{2}\sim b_{2}\sim b_{3}\sim O(1/N)  $ and
$ a_{3}\sim O(1/N^{2}) $. In the same time condition
$ \langle [O_{3}^{c,(1)}](x) [O_{3}^{c,(2)}](y)\rangle =0$ gives
the equation entangling $b_{1}$ and $a_{1}$,
($a_{1}=a_{1}^{0}/N+O(1/N^{2})$
and
$b_{1}=b_{1}^{0}+O(1/N)$):
\begin{equation}
\langle [O_{3}^{(1)}](x) [O_{3}^{(2)}](y)\rangle +
b_{1}\langle [O_{3}^{(1)}](x) [O_{3}^{(1)}](y)\rangle +
a_{1}\langle [O_{3}^{(2)}](x) [O_{3}^{(2)}](y)\rangle=O(N^{-4})
\label{ort}
\end{equation}
To determine the coefficient
$ b_{1}^{0} $ one must use VS scheme described
above~(see also Ref.~\cite{Step}):
$ b_{1}^{0}=
2\gamma_{21}/(\gamma_{O^{(1)}_{3}}-\gamma_{O^{(2)}_{3}}) $;
where
$ \gamma_{{O^{i}_{3}}} $ are the anomalous dimensions of the
corresponding operators in the $1/N$ order:
(All answers it will be given in the units
$(\eta_{1}/N)$ and $(\eta_{1}/N)^{2}$ for the first and second order
 of
$ 1/N $ expansion, respectively.)
\FL
\begin{eqnarray}
\gamma_{O^{(1)}_{3}}&=&
6(2\mu-1)[\mu(2\mu-3)-(\mu-1)]/(2-\mu)\nonumber\\
\gamma_{O^{(2)}_{3}}&=&  \>
(2\mu-1)
\{(2\mu-3)[27\mu-3(11\mu+4)+\mu^2+9\mu+2]-12(\mu-1) \}
/3(2-\mu).
\label{vsdim}
\end{eqnarray}
The coefficient
$ \gamma_{21} $ arises from the mixing of the operators
$O^{(1)}_{3} $  and $O^{(2)}_{3}$ in
$ 1/N $ order:
\FL
\begin{equation}
\gamma_{21}=\lim_{\Delta\to 0}\Delta\sum_{\{ G\}} n^{\sigma}
\Gamma_{G}^{\sigma^{3},2}(p,p_{1},p_{2})
={2\mu(2\mu-3)(4\mu^{2}-1)}/{3(2-\mu)}+O(1/N^{2}).
\label{q_{21}}
\end{equation}
Since  $ b_{1}^{0} $ is known,
$ a_{1}^{0} $ can be easy obtains from the
orthogonality condition~(\ref{ort}). One need calculate
two correlators
$ \langle [O_{3}^{(1(2))}](x) [O_{3}^{(1(2))}](y)\rangle$
in the leading
order and one
$ \langle [O_{3}^{(1)}](x) [O_{3}^{(2)}(y)]\rangle $
in the next to leading order.
Note,  for the determination
of the same coefficient in VS scheme
one need to calculate
$ 20 $ diagrams in
$ 1/N^{2} $  order.

Obviously, the same scheme can be applied for the
determination of coefficient $ \alpha_{1} $ in the
operator $ [O_{s}] $~(see Eq.~\ref{Os}). However, for the
persuasiveness  we have carried out calculations in the
both approaches, which result in the same answer
$ \alpha_{1}=\eta_{1}s(s-1)\bar\alpha_{1} $:
$$ \bar\alpha_{1}=
\frac{(2\mu-3)(\mu-1)}{4(3-\mu)}
\frac{4s(\mu-3)+\mu^2-5\mu+12 }{4s(2\mu-3)-\mu^2-7\mu+12}.
$$
Thus this trick allows to fix the form of the scaling operator
avoiding the cumbersome
$ 1/N^{2} $ order calculations.
(The coefficient
$ \alpha_{1} $ is singular at
$ \mu=\mu_s\simeq3/2(1+1/16(s-1)) $. But this fact has the simple
explanation~--~at this point the degeneration of 1/N order
anomalous dimensions occurs (see e.g. formula for
$ b_{1} ^{0}$). So it is only a artifact of used approach, which has to
be modified in this case.)

Since the coefficient
$ \alpha_{1} $
is known the anomalous dimension of
$ [O_{s}] $
can be determined from Eq.~(\ref{main}) for
$ \Gamma_{s}(p,p_{i})=\langle
[O_{s}](p)\sigma(p_{1})\ldots\sigma(p_{s})\rangle $.
It is instructive to check that the {\sl lhs} and {\sl rhs} of
Eq.~(\ref{main}) have the same momentum dependence. This  can be
done on the formula level and leads us to very nice formula for
$ u_{s}^{(2)} $
($ u^{(2)}_{s}=(-\gamma_{s}^{(2)}+s\gamma_{\sigma}^{(2)}) $) having
obvious resemblance with its counterpart in dimensional
regularization scheme:
\begin{equation}
u^{(2)}_{s}=2\Delta\sum_{G}n^{G}_{\sigma}[KR' G]_{\Delta}.
\label{KR}
\end{equation}
Here the sum runs over all diagrams; the
$ KR' $ operation is the standard operation of the subtractions on
the divergent subgraphs;
$ [..]_{\Delta} $ means that only first order poles in
$ \Delta $ should be picked out;
$ n^{G}_{\sigma} $ is the number of
$ \sigma $  lines in a diagram
$ G $.
It is convenient to represent
$ u_{s}^{(2)} $ as:
\FL
\begin{equation}
u_{s}^{(2)}={s(s-1)}u_{2}^{(2)}/2+s(s-1)(s-2)\left \{
r^{(2)}_{3}+ 2\bar\alpha_{1}\gamma_{21}\right \}.
 \label{u2}
\end{equation}
Here
the first  two terms  arises from
the diagrams describing transition $ \sigma^{2}\to \sigma\sigma $
($u_{2}^{(2)}$),
and
$ \sigma^{3}\to\sigma\sigma\sigma  $  ($r^{(2)}_{3}$) in the
correlator $ \Gamma^{\sigma^{s},s}(p,p_{i}) $,
while the last one is
 due to the admixture in $ [O_{s}] $.
Since the anomalous
dimension of
$ \sigma $ field ($ \gamma_{\sigma} $)
and those of $ \sigma^{2} $
operator~
($ \gamma_{2}=2\gamma_{\sigma}-u_{2} $) are known up to
 $ 1/N^{2} $
order~\cite{VPH,jag,jag2}, the problem reduces to the
determination of $ r^{(2)}_{3} $.
There are 26 diagrams
(three $3$-loop, ten $4$-loop, ten $5$-loop,
 and three $6$-loop) which contribute to
$ r^{(2)}_{3} $. They all can be calculated
with the help of technique developed in~\cite{VPH,VPH2}.
The final expression for
$ r^{(2)}_{3} $  reads:
\widetext
\FL
\begin{eqnarray}
r^{(2)}_{3}&=&
\frac{\mu^2(\mu-1)(15\mu^3-51\mu^2+52\mu-12)}{2(\mu-2)^{2}}C+\nonumber\\
&&\frac{\mu(2\mu-3)(104\mu^5-383\mu^4+542\mu^3-383\mu^2+104\mu-8)}
{12(\mu-1)(\mu-2)^{2}}
\end{eqnarray}
\noindent
Here
$ C=\psi'(1)-\psi'(\mu-1) $ and
$ \psi(x)=\Gamma'(x)/\Gamma(x)$.
We do not give the complete expression for the critical exponents
$ x_{s} $ because of its large size.
It is more interesting to look on the corresponding expansions
near $2$ and $4$ dimensions
(we write down only leading in
$ s $ terms):
\FL
\begin{mathletters}
\label{ex24}
\begin{eqnarray}
x_{s}^{2+\epsilon}&=&2s-s(s\epsilon/N)-s/2(s\epsilon/N)^{2}+..
\label{2pe}\\
x_{s}^{4-\epsilon}&=&2s+6s(s\epsilon/N)-34s(s\epsilon/N)^{2}+..
\label{4me}
\end{eqnarray}
\end{mathletters}
It should be stressed that the coefficient
at the last term in Eq.~(\ref{2pe}) differs from those
obtained in the paper~\cite{CC}. In the
$ 4-\epsilon $ expansion the exponents
$ x_{s} $ (corresponding to
$ (\phi^{2})^{s} $ operators) has been so far known with
$ \epsilon $ accuracy only. For the additional check of our results
we have calculated
$ x_{s} $ up to
$ \epsilon^{2} $ order:
\FL
\begin{eqnarray}
x_{s}&=&2s+6\epsilon\frac{s(s-2)}{N+8}-\epsilon^{2}s\left\{(s-1)
[34(s-2)(N+8)+
\right.
\nonumber \\&& \left.
(11N^{2}+
 92N+212)]
-(13N+44)(N+2)/2)\right\}/{(N+8)^{3}},
\end{eqnarray}
the above expression  being in the full agreement with those obtained in
the frame of 1/N expansion.
\vskip 0.5cm

Let us now discuss the obtained results. One can see from the
Eqs.~(\ref{ex24}) that as in the
$ 2+\epsilon $ as  in the
$ 4-\epsilon $ expansions of the exponents
$ x_{s}$  the second order terms have negative
 ("wrong") sign. Thus for  a large
$ (s\epsilon/N) $  the critical exponents given by the truncated
series~(\ref{ex24}) become a negative. This might serve a starting point
for the speculations on the stability of the conventional fixed
points. But, do the first order terms~(\ref{ex24}) give a well
approximation for
$ x_{s} $ when $ (s\epsilon/N) $ is large?
The answer, of course, is negative, since
the estimative terms in~(\ref{ex24})
are
$ O((s\epsilon/N)^{3})$ order.
Moreover,
 the pure combinatorical analysis shows
that   the
$ k $-th order term in the series~(\ref{ex24})
behaves as $s^{k+1}$ at  large
$ s $. Thus to obtain the answer for
$ x_{s} $ which were sensible for large
$ s\epsilon/N $ one need sum up all
order corrections~(see for further discussion Refs.~\cite{HB,DKM}).
Even taking into account the recent  progress in the higher
order calculations~\cite{jag2} the feasibility
 of
this program  causes the great doubts.

The said above concerns both the
$ 2+\epsilon, \> 4-\epsilon $ expansions and the
$ 1/N $ expansion.  However, in the case when only a few first terms
are calculable, the
$ 1/N $ expansion has the definite advantage in comparison with
$ \epsilon $  ones. Indeed,
$ 1/N $ expansion is more informative, because it contains information
on  critical exponents in whole interval
$2<\mu<4$.
In the case under consideration, one can,
with some extent of a confidence, judge about the general tendency
from the first order results.

\begin{center}
\begin{figure}
\centerline{\psfig{figure=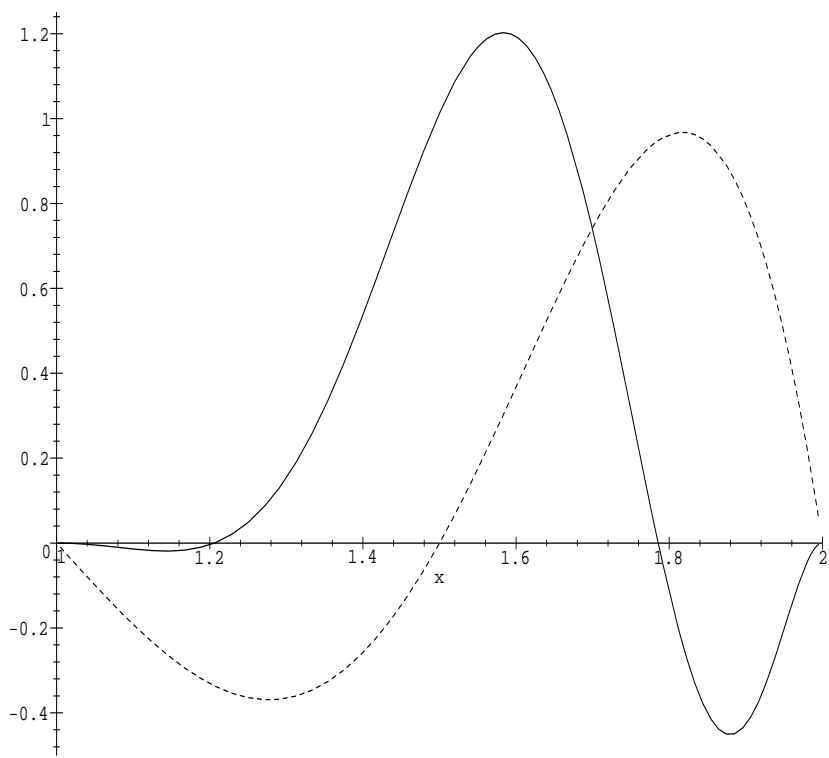,height=14cm,width=14cm,angle=-90}}
\caption{ The functions  $ A_{1}(x) $  and $ A_{2}(x) $
are plotted as the functions of the space dimension.
The dot line corresponds to  $ A_{1}(x) $
(the first order in $ 1/N $),
and the solid line~---~to
$ A_{2}(x) $
(the second order in $ 1/N $).}
\label{reduced}
\end{figure}
\end{center}

Let the function
$ A_{k}(\mu) $ is the coefficient at the leading in
$ s $  term
($ s^{k+1} $) of the
$k $ order term in the
$ 1/N $ expansion of the exponent
$ x_{s} $.  The function
$ A_{1}(\mu) $ and
$ A_{2}(\mu) $ are drawn in Fig.~\ref{reduced}.
It is natural to consider the
intervals
$ I_{k} $ on which these functions are negative, i.e. if
$ \mu\in I_{k} $  then
$ A_{i}\leq 0 $ for
$ i=1,\ldots,k $. (Indeed, it might be said that the stability
problem in the
$ 2+\epsilon  $ expansion arises from the negativity of the first two
coefficients in the expansion~(\ref{2pe}). If one of them were
positive, the problem would hardly be considered as serious one.)
It is seen that the interval
$ I_{2} $ is considerably smaller than
$ I_{1} $ (
$ I_{1}=[1,\>1.5]$, while  $I_{2}\simeq[1,\> 1.2]$).
Obviously, if this tendency
($ I_{k}\to 0 $ at
$ k\to\infty $) will hold in the higher order, the stability problem
lost its sharpness.
 Indeed, in this
case for any $ \epsilon>0 $ the $ 1/N $ corrections starting from
some order will shift the critical dimensions in the
true~--~irrelevant~--~direction
(at least they will be sign~-~varied).

We conclude with the following remarks:
In all available expansions of the
$ N $ vector model ($ 2+\epsilon,\> 1/N,\> 4-\epsilon $)
there exists the class of
operators which  acquire a "large" anomalous dimensions in the first
orders of the corresponding expansions. This fact is not related to
the expansion used, but has the combinatorical origin. The stability
problem arises when one tries to extrapolate the first orders results
out  of their range of applicability. Though we do not state the
$ 2+\epsilon $  expansion is  free of the problems~\cite{CH,KZ}
it seems they are not related to the high~--~gradient operators.

\acknowledgments
The authors are grateful to Prof. A.~N.~Vasil'ev
and Dr. S.~Kehrein for the valuable
discussions.
This work was supported by Grant 97--01--01152 of the
Russian Fond for Fundamental Research and by INTAS Grant
93--2492--ext and was carried out under the research program of the
International Center for Fundamental Physics in Moscow.

\begin{references}
\bibitem{KLY}
V.E. Kravtsov, I.V. Lerner and V.I. Yudson,\
Zh. Eksp. Teor. Fiz. {\bf 94}, 255  (1988);
Sov. Phys. JETP {\bf 67}, 1441 (1988);
Phys. Lett. {\bf A 134}, 245 (1989).
\bibitem{WZ}
F. Wegner,\  Z. Phys. {\bf B 78}, 33 (1990).
\bibitem{W1}
F. Wegner,\  Nucl. Phys. {\bf B[FS] 354}, 141 (1991).
\bibitem{W2}
H. Mall and F. Wegner,\  Nucl. Phys. {\bf B[FS] 393}, 495 (1993).
\bibitem{Step}
A.N. Vasil'ev and A.S. Stepanenko,\  Teor. Mat. Fiz. {\bf 95}, 160 (1993);
Theor. Math. Phys. {\bf 94}, 471 (1993).
\bibitem{Ruhl}
K. Lang and W. R\"uhl,\ Nucl. Phys. {\bf B[FS] 400}, 597 (1993).
\bibitem{CC}
G.E. Castilla and S. Chakravarty,\  Phys. Rev. Lett. {\bf 71}, 384 (1993).
\bibitem{CC1}
G.E. Castilla and S. Chakravarty,\  Nucl.Phys. {\bf B[FS] 485}, 613 (1997).
\bibitem{VPH}
A.N. Vasil'ev, Yu.M. Pis'mak and J.R. Honkonen,\ Theor. Math. Phys.
{\bf 47} 465 (1981)
\bibitem{VPH2}
A.N. Vasil'ev, Yu.M. Pis'mak and J.R. Honkonen,\ Theor. Math. Phys.
{\bf 50} 127 (1982)
\bibitem{jag}
J.A. Gracey,\
Progress with large $N_{f}$ $\beta$-functions, Preprint
{\bf {LTH 383}}, (hep-th/9609409)
\bibitem{jag2}
D.J. Broadhurst, J.A. Gracey, D. Kreimer\
Beyond the triangle and uniqueness relation:
non-zeta counterterms at large
$ N $ from positive knots, (hep-th/9607174)
\bibitem{Nal}
A.N. Vasil'ev and M.Yu. Nalimov,\ Teor. Mat. Fiz. {\bf 55}, 163 (1983);
Theor. Math. Phys. {\bf 55}, 423 (1983).
\bibitem{Osborn}
H. Osborn and A. Petkou,\ Ann.Phys.{\bf \underline{231}} 311 (1994);
\bibitem{Petkou}
A. Petkou,\ Ann.Phys.{\bf \underline{249}} 180 (1996);
\bibitem{Pol}
A.M. Polyakov,\ JETF, {\bf 66} 23 (1974);
\bibitem{HB}
E. Brezin  and S. Hikami,\ {\em Fancy and facts in the ($ d $-2)
expansion of non-linear sigma models}; cond-mat/9612016;
\bibitem{DKM}
S. Derkachov, S.~Kehrein and A.N. Manashov, {\em High gradient
operators in the N~-~vector model}; to be published in Nucl.Phys.B;
\bibitem{CH}
J.L. Cardy and H.W. Hamber, Phys. Rev. Lett. {\bf 45} 499 (1980);
\bibitem{KZ}
H. Kunz and G. Zumbach, J. Phys. A: Math. Gen. {\bf 22} L1043 (1989);
\end{references}
\end{document}